\documentclass[aps,preprint,showpacs]{revtex4}
\usepackage{graphicx,epsfig}
\usepackage{amssymb}
\begin{document}

\title{ Redshift of light emitted by particles orbiting a black hole immersed in a strong magnetic field}

\author{L. A. L\'opez $^{1}$ }
\email{lalopez@uaeh.edu.mx}
\author{Nora Bret\'on $^2$}
\email{nora@fis.cinvestav.mx}

\affiliation{$^1$  \'Area Acad\'emica de Matem\'aticas y F\'isica.,  UAEH, carretera Pachuca-Tulancingo km 4.5, C.P. 42184, Mineral de la Reforma, Hidalgo, M\'exico}
\affiliation{$^2$ Dpto de F\'isica, Centro de Investigaci\'on y de Estudios Avanzados del I.P.N, Cinvestav, 
A. P.  14-740, Mexico City, Mexico}

\begin{abstract}
In this paper we analyze the frequency shifts of the light emitted by particles describing stable circular geodesics around a static black hole immersed in an external magnetic field of arbitrary strength. This system is represented by the Ernst solution of the Einstein-Maxwell equations. The presence of the magnetic field and its magnitude affects both the geodesics and the red-blueshifts of the light emitted by  neutral or charged particles orbiting the black hole. When the magnetic field is turned off we recover the characteristic redshifts coming from particles orbiting a Schwarzschild black hole. 

\end{abstract}

\pacs{04.70.Bw,04.70.-s, 04.30.Nk}

\maketitle

\section{Introduction}

Astrophysical objects such as black holes have been of great interest to the scientific community for many years; this interest has increased even more due to the possibility of observing strong field gravitational phenomena. For example the analysis of black hole stability played and outmost role in characterizing gravitational-wave signals detected \cite{Abbott2016}.
The project GRAVITY \cite{2009ASSP}  tracks the stars orbiting around the supermassive black hole at the center of our galaxy and the Event Horizon Telescope (EHT) collaboration was able to obtain an image of the supermassive black hole in M87* \cite{Akiyama2019}, which opens the way for a phenomenological approach for a deeper understanding of Black Holes (BH). 
 
Therefore, it is necessary to study the relationship between the observed gravitational phenomena and the parameters that characterize BH, such as the mass, angular momentum or charge. In this sense in \cite{Cardoso2009} was shown the relationship among parameters of the circular null geodesics, Lyapunov exponents, and quasinormal modes of black holes (BH). In this context, different BH solutions have been studied \cite{Fernando2012} \cite{Breton2016} \cite{Lopez2018}.In \cite{Konoplya2017} the limits of applicability of this correspondence were addressed.

Also in \cite{Herrera-Aguilar2015} the authors developed a theoretical approach to obtain the parameters of a Kerr BH in terms of the redshift and blueshift of photons emitted by particles traveling along stable circular geodesics. Using this idea, different BH configurations have been studied; in \cite{Becerril2016} were derived the red/blueshifts of static spherically symmetric BH and in \cite{Kraniotis2021} were studied  the red/blueshifts and  frame dragging of the 
Kerr–Newman–de Sitter and Kerr–Newman  BH.
On the other hand, observational evidence indicates that in the center of each galaxy there are black holes \cite{Begelman2003} and always accompanied by magnetic fields whose origin may be  external or  generated by currents in the accretion disk.

The Ernst solution  \cite{Ernst1976} of the coupled Einstein-Maxwell equations describes the gravity of a static black hole immersed in a uniform magnetic field; it is also known as  Schwarzschild–Melvin (SM) BH. Despite being non asymptotically flat, the magnetic Ernst solution is a useful model for a black hole in certain astrophysical situations. For charged particles geodesics of  the Ernst BH have been described in \cite{Hoenselaers1979}  as well as for Melvin Universe  (magnetic universe), obtaining that for charged particles bound orbits always exist for realistic magnetic field strengths.

Frequently it has been considered the external magnetic field as a perturbation or test field (test field approximation); for instance, the Wald \cite{Wald1974} solution consists in a test magnetic field that does not affect the curvature. This solution has been addressed in \cite{Frolov2010}, where the bounded trajectories of charged particles near a weakly magnetized Schwarzschild BH were derived; in this setting neutral particles are not affected by the presence of the weak magnetic field. In \cite{Lim2015} is studied the influence of an external uniform electric or magnetic field on charged particles via the Lorentz interaction. In \cite{Tursonov2016} the analysis of circular orbits and related quasi-harmonic oscillatory motion of charged particles around weakly magnetized rotating black holes was addressed. In \cite{Hackmann2020} were studied the Innermost Stable Circular Orbits (ISCOs) of charged particles in a weak electromagnetic field in the vicinity of a Schwarzschild BH.

In \cite{Aliev2002} the motion of charged particles around a rotating BH in a weak magnetic field was analysed, obtaining that the presence of the magnetic field enlarges the region of  marginal stability shifting the radius of the marginal stable orbits towards the  horizon. In \cite{Konoplya2007}  corrections to the bending angle and time delay
due-to presence of weak magnetic fields in galaxies were estimated.

As was briefly summarized, charged particles trajectories have been thoroughly studied in the spacetime of a BH in a weak magnetic field.  Most of the treatments apply to test magnetic fields that do not alter curvature, therefore it has not effect on uncharged or neutral particles. The advantage of studying an exact solution of the Einstein-Maxwell equations is that the magnitude of the magnetic field is arbitrary and, as we show in the following,   acting through curvature it has an effect even on neutral test particles, enlarging as well the region of stable circular orbits by pushing the ISCO orbits towards the horizon.

In this work our aim is to determine the redshift of the light emitted by particles orbiting the Ernst BH in stable circular trajectories; therefore we focus on these kind of orbits for both, charged and uncharged test particles, pointing out that the latter are indeed influenced by the magnetic field even if it is not too strong; we  present as well the region of the ISCO, in terms of the dimensionless parameter $mB$  ($m$ the BH mass and $B$ the magnetic field). Moreover, we determine the  upper bound on $mB$ that allows stable circular orbits.Then is determined  the influence of the external magnetic field on the redshifts coming from particles (charged or uncharged) orbiting in stable circular trajectories around the Ernst BH.

The paper is organized as follows: in Section II the  Ernst or Schwarzschild–Melvin (SM) BH is introduced as well as the effective potentials for charged and uncharged particles. In  section III  a short summary is given to determine the  frequency shifts of photons emitted from particles  moving in stable geodesics around a static BH, also the frequency shifts of light emitted by neutral and charged particles orbiting the Ernst BH are determined. Finally, conclusions are given in the last section.
\section{The Ernst or Schwarzschild-Melvin black hole.}

The Ernst solution or Schwarzschild–Melvin (SM) BH,  also known as electrified/magnetized Schwarzschild BH, describes the spacetime  of a static BH immersed in an external uniform magnetic or electric field; in spherical coordinates the metric is described by \cite{Ernst1976}, 

\begin{equation}\label{SM}
ds^{2}=\Lambda^{2}(-\Delta dt^{2}+\Delta^{-1}dr^{2}+r^{2}d\theta^{2})+\Lambda^{-2}r^{2}\sin^{2} \theta d\phi^{2},
\end{equation}

with $\Lambda=1+\frac{1}{4} B^{2}r^{2} \sin^{2} \theta$ and $\Delta=1-\frac{2m}{r}$, where $m$ is the BH mass  and $B$ is the external  magnetic field parameter. 
As was shown in \cite{Walter1980}  where the Gaussian curvature was examined, the event horizon remains the same as for Schwarzschild, located at $r = 2m$ and the curvature singularity at  $r=0$. When $m \rightarrow 0$ the metric reduces to the Melvin's magnetic universe \cite{Melvin1965}.

It is worth to note that the effect of the magnetic field resembles the one of a cosmological constant: writing the $tt$ metric component (with $\theta= \pi/2$, just for simplicity), 

\begin{equation}
g_{tt}= - \Delta \Lambda^2= - \left( 1 + \frac{B^2 r^2}{2} +  \frac{B^4 r^4}{16} - \frac{2m}{r} \Lambda^2 \right),  
\end{equation} 
the second term with dependence on $r^2$ acts similarly to a positive cosmological constant, from which we can guess the confining effect that the magnetic field exerts on test particles as well as on light.
Also due to the presence of the electromagnetic field, the metric (\ref{SM}) is not asymptotically flat.

The vector potential  $ A_{\mu} = (A_t,0,0, A_{\phi})$, for an electric field $\mathcal{E}$ and magnetic field $B$,   is given by \cite{Lim2015}, \cite{Konoplya2008}, 

\begin{equation}\label{Ap}
A_{t}=  \mathcal{E} r  \Delta \Lambda \cos{\theta}, \quad  A_{\phi}= \frac{1}{2 \Lambda}Br^{2}\sin^{2}\theta.
\end{equation}

In the context of astrophysics typically magnetic fields  are considered in the test field regime, that do not influence the spacetime curvature.  In that case the metric is simply the Schwarzschild metric ($\Lambda=1$ in (\ref{SM}) ) with the  magnetic field associated to $A_{\phi}$.

\subsection{ Effective Potential for a charged test particle}

The equations of motion of a charged particle with mass $m_{c}$ and charge $q_{c}$ may be derived from the Lagrangian; 

\begin{equation}\label{Lagq}
\mathfrak{L}=\frac{1}{2}g_{\mu \nu}\dot{x}^{\nu}\dot{x}^{\mu}+ \kappa A_{\mu}\dot{x}^{\mu},
\end{equation}

where $\kappa = q_{c}/m_{c}$ is the  specific charge  of the test particle and $A_{\mu}$ is the electromagnetic potential. 

The momenta conjugate is given by

\begin{equation}\label{P}
P_{\mu}=g_{\mu \nu}\dot{x}^{\nu}+\kappa A_{\mu}.
\end{equation}

For  axisymmetric stationary spacetimes a test particle has two conserved quantities, 
its energy and its angular momentum, related to the two Killing vectors $\partial_{t}$ and $\partial_{\phi}$, respectively,

\begin{equation}\label{Ec}
P_{t}=g_{tt}\dot{t}+\kappa A_{t}=g_{tt}{U}^{t}+\kappa A_{t}=-{E},  
\end{equation} 
and
\begin{equation}\label{Lc}
P_{\phi}=g_{\phi\phi}\dot{\phi}+\kappa A_{\phi}=g_{\phi\phi} {U}^{\phi}+\kappa A_{\phi}= {L}. 
\end{equation} 

The components of the $4-$velocity of the test particle are ${U}^{t}=-{({E}  +\kappa A_{t})}/{g_{tt}}$ and
${U}^{\phi}= {({L}-\kappa A_{\phi} )}/{g_{\phi\phi}}$. If the $4-$velocity is normalized to unity ${U}^{\mu} {U}_{\mu}=-1=g_{tt}({U}^{t})^{2}+g_{rr}({U}^{r})^{2}+g_{\theta \theta}({U}^{\theta})^{2}+g_{\phi \phi}({U}^{\phi})^{2}$ then we obtain;

\begin{equation}
-1=\frac{({E}+\kappa A_{t})^{2}}{g_{tt}}+g_{rr}\dot{r}^{2}+\frac{({L}-\kappa A_{\phi})^{2}}{g_{\phi \phi}}+g_{\theta\theta}\dot{\theta}^{2}.
\end{equation}

\begin{figure}[h]
\begin{center}
\includegraphics [width =0.45 \textwidth ]{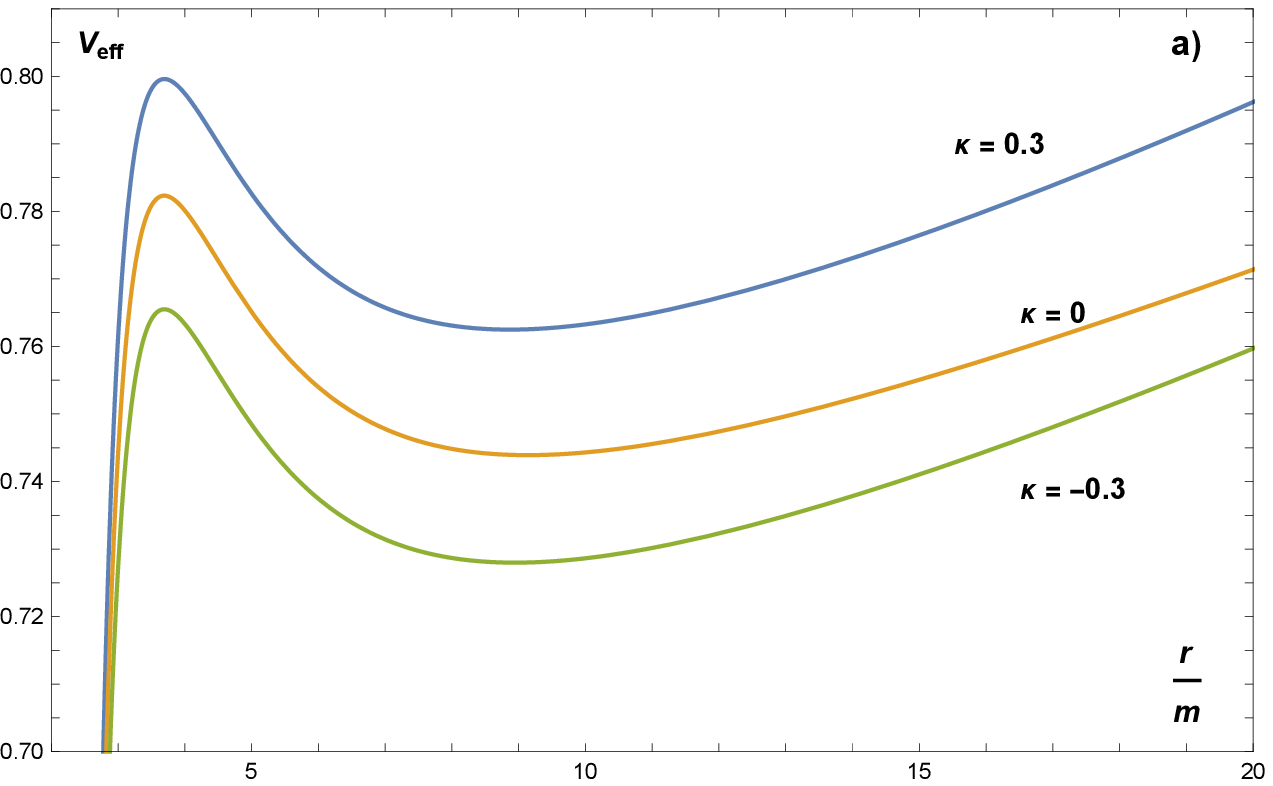}
\includegraphics [width =0.45 \textwidth ]{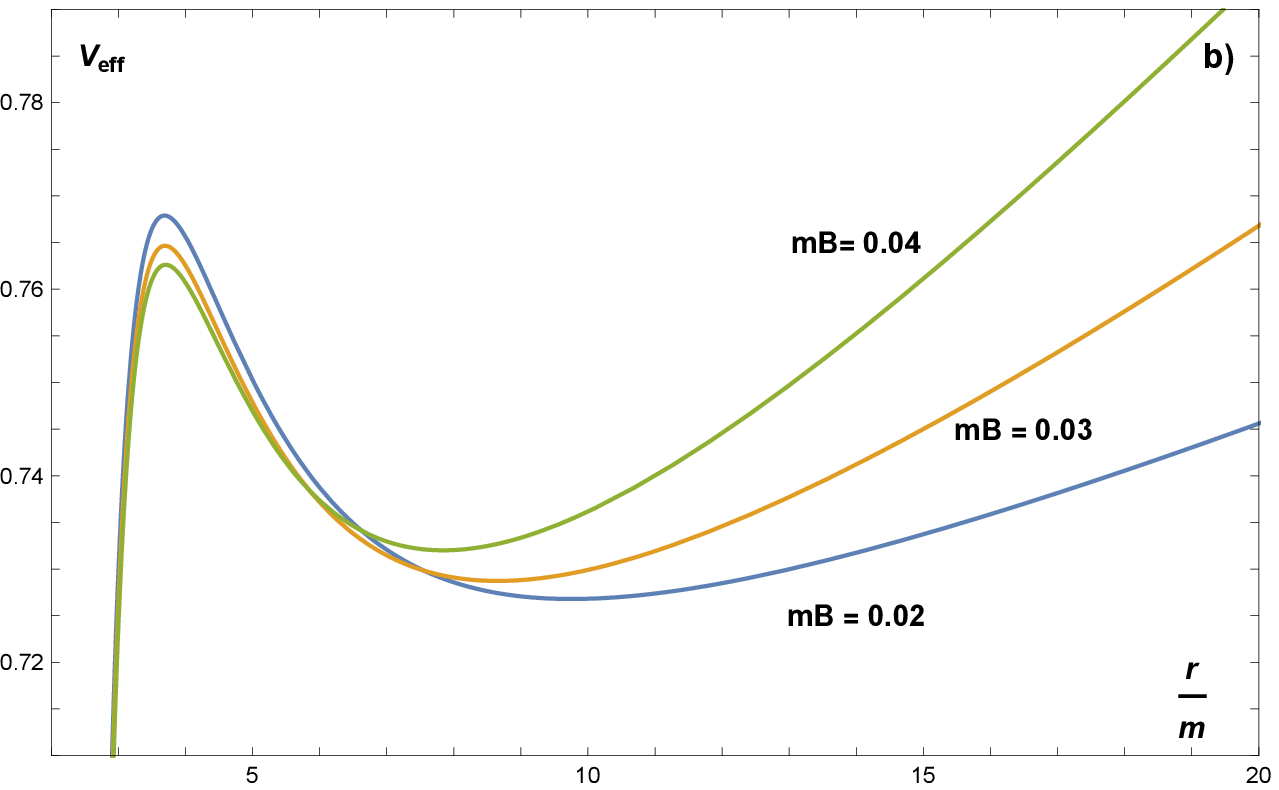}
\end{center}
\caption{a) The behavior of the effective potential ${V_{\rm eff}}$ for different values of the specific charge of the test particle, $\kappa$, with $E=0.5$, $\frac{L}{m}=2.1$ and  $mB=0.027$. b) The behavior of the effective potential ${V_{\rm eff}}$ varying $mB$ with $\frac{L}{m}=2.1$, $E=0.5$ and $\kappa=0.3$ The effect of the magnetic field is of confining, even for uncharged test particles ($\kappa = 0$).}
\label{Vefq}
\end{figure}

Comparing with $g_{rr}\dot{r}^{2}+{V_{\rm eff}}=0$ we get the effective potential over the test particle, which depends on ${E}$ and ${L}$,

\begin{equation}\label{Vq}
{V}_{\rm eff}=1+ \frac{({E}+\kappa A_{t})^{2}}{g_{tt}}+\frac{({L}-\kappa A_{\phi})^{2}}{g_{\phi \phi}}.
\end{equation}
We shall consider only magnetic field, i.e. $\mathcal{E}=0, A_t=0$; for equatorial orbits $\theta=\pi/2$ and using (\ref{SM}) and the vector potential for the magnetic field (\ref{Ap}) we obtain that

\begin{equation}\label{Vq1}
{V_{\rm eff}}=1-\frac{ {E}^{2}}{\Delta \Lambda^2}+\frac{{\Lambda}^{2}}{r^{2}}\left( L- \kappa \frac{B r^{2}}{2 \Lambda}\right)^{2}.
\end{equation}
In Fig. \ref{Vefq}\textbf{ a)} is shown the behavior of ${V_{\rm eff}}$, Eq.  (\ref{Vq1}), for different values of the specific charge $\kappa$ and Fig. \ref{Vefq} \text\bf{b)} shows ${V_{\rm eff}}$ for different values of the dimensionless parameter $mB$. The effective potential presents  maximum and minimum  that indicates there exist circular orbits, both, unstable and stable.
Therefore when considering a charged particle with mass $m_{c}$ and charge $q_{c}$,  the effect of the effective potential depends on the magnitude and sign of the specific charge $\kappa$. When $mB$ increases, the confining effect increases, as can be seen in Fig. \ref{Vefq} \textbf{b)}.

When we consider the equation for radial motion $g_{rr}\dot{r}^{2}+{V_{\rm eff}}=0$, the circular orbits correspond to the radii $r_{c}$, where the potential and its derivative are zero ( $V_{\rm eff}(r_c)=0$ and ${V_{\rm eff}}^{'}(r_{c})=0$). Then for circular orbits, from Eq.  (\ref{Vq1}) the following expression, that restricts $L$, should be fulfilled

\begin{equation}\label{Lk}
\left({L}-\kappa A_{\phi}\right)^{2}a(r_{c}) +\left({L}-\kappa A_{\phi}\right), b(r_{c})+c(r_{c})=0
\end{equation} 

where;

\begin{equation}
a(r_{c}) =\frac{(4+B^{2}r^{2})}{8(r-2m)r^{3}}[B^{2}r^{2}(3r-5m)-4(r-3m)],
\end{equation}

\begin{equation}
b(r_{c})=\frac{4B\kappa}{r}, \quad c(r_{c})=\frac{2[4m+B^{2}r^{2}(2r-3m)]}{r(r-2m)(4+B^{2}r^{2})}.
\end{equation}

\begin{figure}[h]
\begin{center}
\includegraphics [width =0.38 \textwidth ]{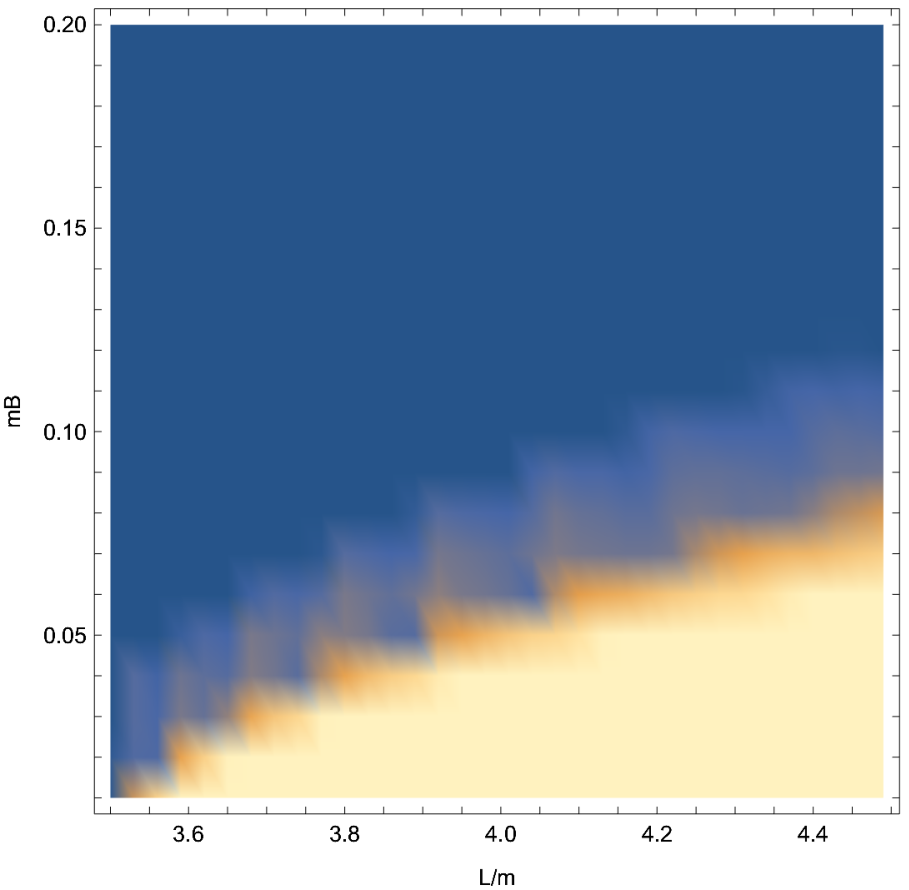}
\includegraphics [width =0.38 \textwidth ]{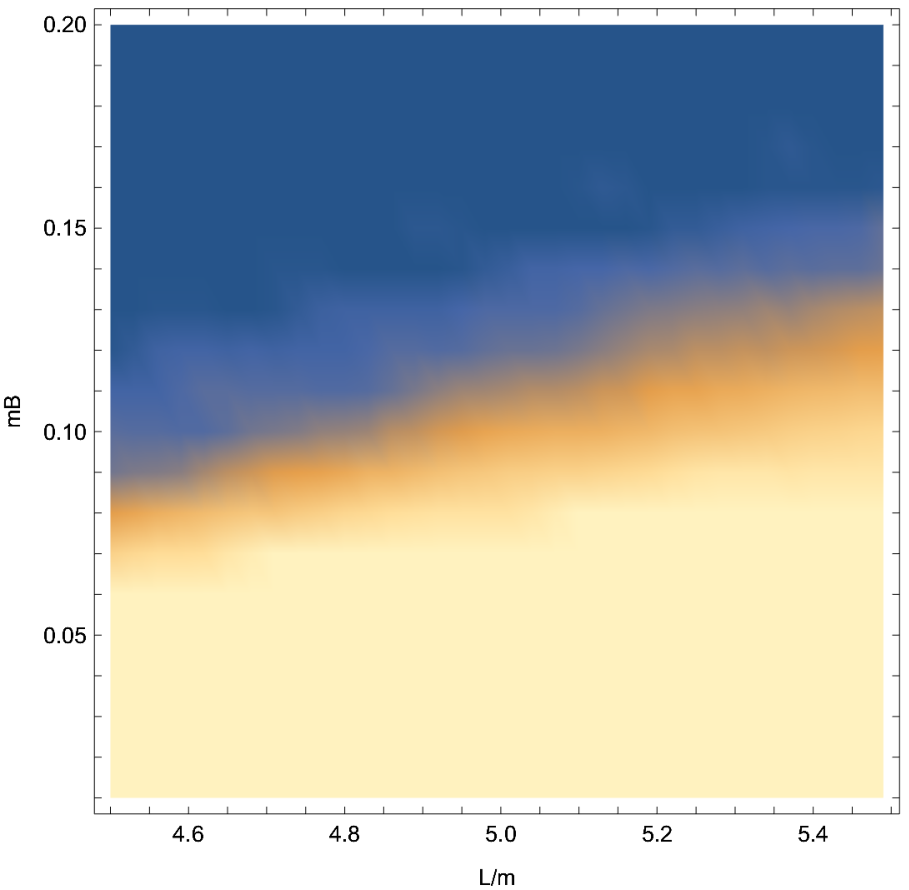}
\includegraphics [width =0.38 \textwidth ]{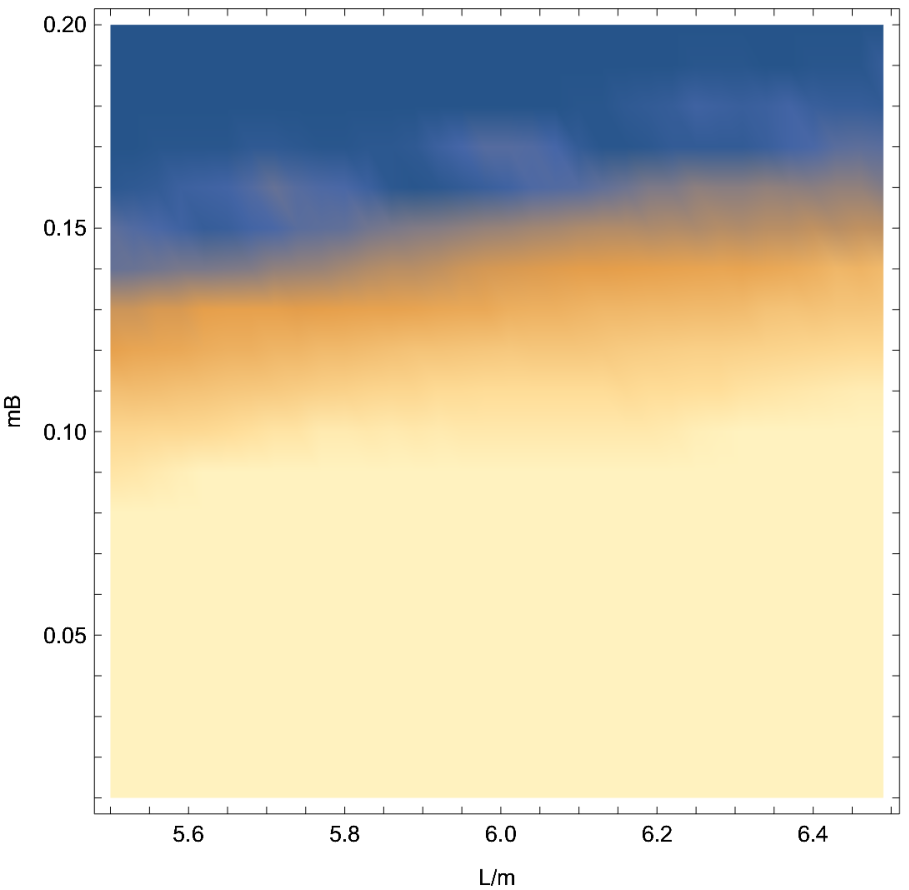}
\includegraphics [width =0.38 \textwidth ]{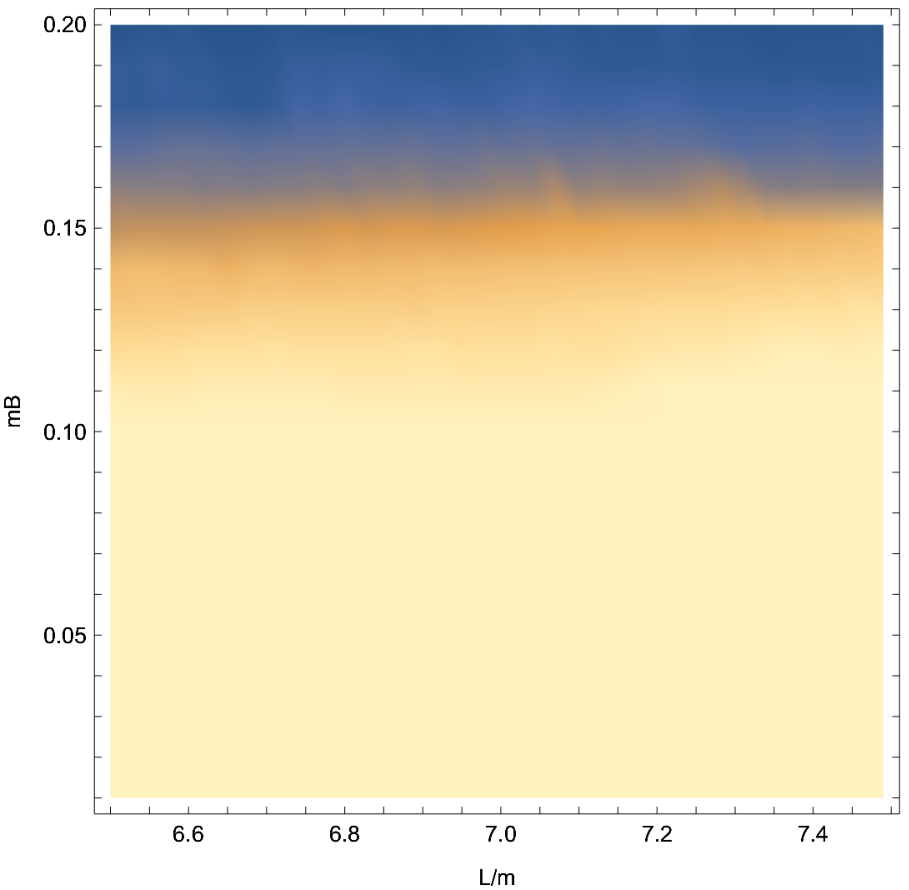}
\end{center}
\caption{Density plots for the parameters ${mB}$ and ${L}/m$; the lighter regions correspond to the pairs ($L/m,B$) of the stable circular orbits of the Ernst BH;  the ranges are $0 \leq mB \leq 0.2$ and $0 \leq \kappa \leq 1$.}
\label{DE}
\end{figure}

Moreover, the stability of the circular orbits requires that $V_{\rm eff}^{''}(r_{c})>0$. This analysis is performed numerically in the range of  $mB$ shown in Fig. \ref{DE}; the density plots for the pairs (${L}/m, {mB}$) correspond to the stable circular orbits; in the ranges $0 \leq mB \leq 0.2$ and $0 \leq \kappa \leq 1$ a large number of stable circular orbits (represented by the lighter part of the graphs) is allowed.
\subsection{Uncharged particles}
The effective potential, Eq.  (\ref{Vq1}),  acting on uncharged test particles, $\kappa=0$ is given by

\begin{equation}\label{EV}
V_{\rm eff}=1-\frac{16rE^{2}}{(r-2m)(4+B^{2}r^{2})^{2}}+\frac{L^{2}}{16 r^{2}}\left( {4+B^{2}r^{2}}\right)^{2}.
\end{equation}

\begin{figure}[h]
\begin{center}
\includegraphics [width =0.45 \textwidth ]{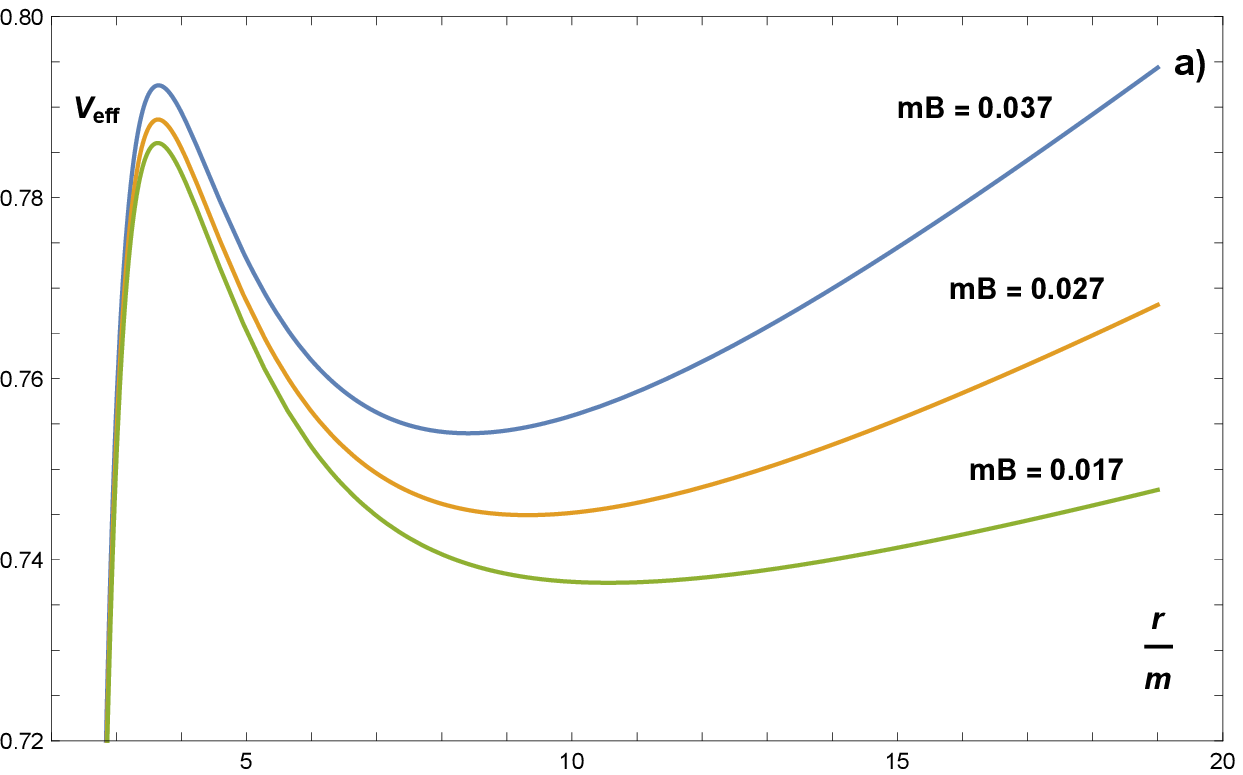}
\includegraphics [width =0.45 \textwidth ]{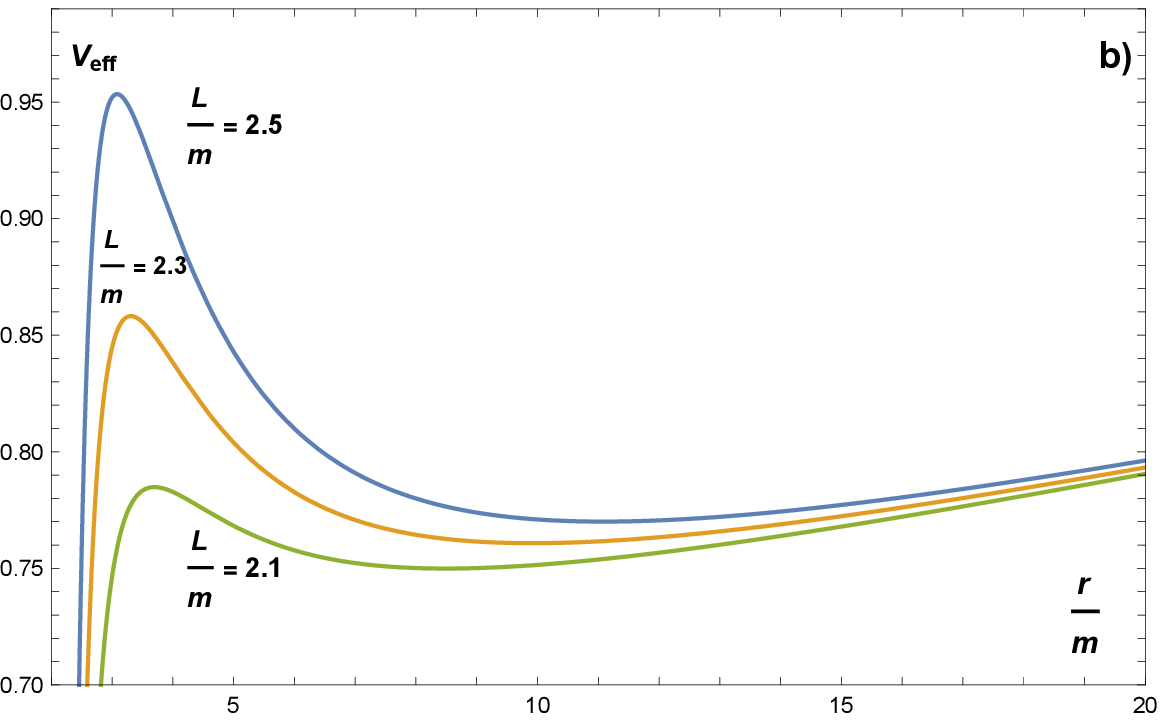}
\end{center}
\caption{\textbf{a)} For uncharged test particles,   the effective potential as a function of $r/m$ is plotted for different values of $mB$ with $E=0.5$ and $\frac{L}{m}=2.1$. \textbf{b)}. The same for different values of $\frac{L}{m}$ with $E=0.5$ and $mB=0.034$.  }
\label{Fig1}
\end{figure}

As shown in Fig. \ref{Fig1}  the effective potential  $V_{\rm eff}$ presents maximum and minimum, corresponding to circular orbits, unstable and  stable, respectively. In Fig. \ref{Fig1}  \textbf{a)}  $V_{\rm eff}$ (\ref{EV}) is shown as a function of $r/m$ for different values of $mB$  and in Fig. \ref{Fig1} \textbf{b)} for different values of  $L/m$. These plots show that uncharged particles are affected by the presence of the external magnetic field, even if it is not too strong, fact that has been disregarded in most of the literature.
If $mB$ increases then the confining effect  increases (see Fig. \ref{Fig1} \textbf{a)}). The same happens when $L / m$ increases (see Fig. \ref{Fig1}\textbf{ b)}) and the maxima of the effective potential is shifted towards the horizon, the effect is the opposite in the minima. 

For the  circular orbits the values of the energy and angular momentum of the test uncharged particle in terms of the BH parameters $(m, B)$ are,  

\begin{equation}\label{energy}
E^{2}\mid_{r_{c}}=\frac{(r-2m)^{2}(4 - r^{2}B^{2})(r^{2}B^{2}+4)^{2}}{16r[4(r-3m)-B^{2}r^{2}(3r-5m)]},
\end{equation}

\begin{equation}\label{angular}
L^{2}\mid_{r_{c}}=\frac{16r^{2}[B^{2}r^{2}(2r-3m)+ 4 m]}{(4+B^{2}r^{2})^{2}[4(r-3m)-B^{2}r^{2}(3r-5m)]},
\end{equation}
evaluated at  $r_c$, the radius of the  circular orbit. 
 
\begin{figure}[h]
\begin{center}
\includegraphics [width =0.5 \textwidth ]{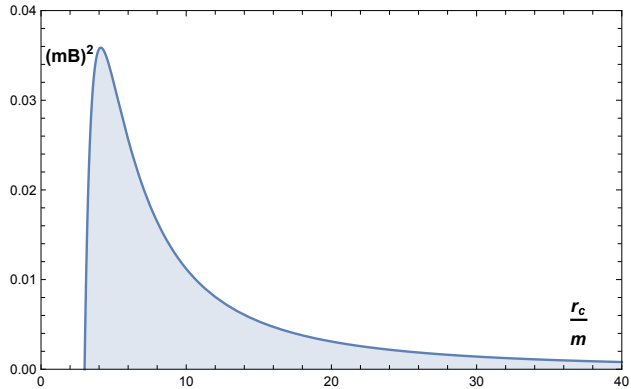}
\end{center}
\caption{The shaded region corresponds to the pairs ($r_{c}, mB$) that define a circular orbit. For values $Bm> 0.189366$, the maximum (not shown in the plot), no circular orbits occur.}
\label{Fig2}
\end{figure}

The conditions $0 \leqslant E^{2}$ and $0 \leqslant L^{2}$ at $r=r_{c}$ lead to the following constrictions between $m$, $B$ and $r_c$:

\begin{equation}
B^{2} < \frac{4(r_{c}-3m)}{r_{c}^{2}(3r_{c}-5m) }, \quad 3m \leqslant r_{c}. 
\label{condB2}
\end{equation}
The second condition is no other than $r_c$ should be larger than the one corresponding to the photosphere radius for Schwarzschild, $r_{ph}^{\rm S}=3m$.  The condition that $r_{c}$  corresponds to a circular orbit that is stable is that the second derivative of the effective potential be positive,  $V_{\rm eff}^{''}(r_c) >0$. In Fig. \ref{Fig2} is shown the region of pairs ($r_{c}/m, (mB)^2$) that correspond to circular orbits. As $mB$ decreases the range of $r_{c}/m$ augments; $mB$ presents a maximum at $mB=0.189366$, this means that for  fields such that $mB > 0.189366$ no circular orbits occur. As $mB$ grows the available range for $r_{c}/m$ is shorter.

As shown in Fig. \ref{Fig2},  the larger number of circular orbits is for values of $mB$  well below the maximum $(mB)_{\rm max}=0.189366$. For concreteness we include  Table  \ref{table1} with the orders of magnitude of the parameter $mB$ and the corresponding BH masses in solar mass units and the magnetic field $B_{\rm SI}$ in Tesla.  To obtain the magnetic field in inverse length units, we use the factor $\sqrt{G \epsilon_{0}}/c^2$, where $G$ is the gravitational constant and $\epsilon_0$ is the vacuum dielectic contant; and $M$, the BH mass,  given in Kg. is transformed to length units with $m= GM/c^2$.
The maximum $(mB)_{\rm max}=0.189366$ could correspond to a ten solar mass BH with a magnetic field of $B_{\rm SI} \approx 10^{16}$ Tesla  (that is huge), or to a more massive BH in a less intense magnetic field.
\begin{table}[!hbt]
    \centering
    \begin{tabular}{|c|c|c|}
    \hline
         BH mass& $B_{\rm SI}$  & $mB$\\
     \hline
        10 & $10^{4}$ & $1.4 \times 10^{-11}$\\
       $10^{6}$  & 1 & $1.2 \times 10^{-10}$ \\
       $10^{6}$ & $10^{8}$ & $ 10^{-2}$\\
       $10^{1}$  & $10^{13}$ & $ 10^{-2}$ \\
    $10^{4}$  & $10^{10}$ & $ 10^{-2}$ \\
    $10^{1}$  & $1.6 \times 10^{16}$ & $ 0.189$\\
    \hline
    \end{tabular}
    \caption{The BH mass in units of solar masses, the magnetic field $B_{SI}$ in Tesla and the corresponding value of the dimensionless parameter $mB$. The first two rows are according to \cite{Piotr2010}. The orders of magnitude we use in this paper are of $10^{-2}$ (three following rows), and finally $(mB)_{\rm max}=0.189366$ could correspond to a ten solar masses BH with a magnetic field of $B_{\rm SI} \approx 10^{16}$ Tesla, or to a more massive BH in a less intense magnetic field. }
    \label{table1}
\end{table}

The condition   $V_{\rm eff}^{''}(r_c) >0$  (stable circular orbits) sets additional bounds on the range of $B$ already restricted by (\ref{condB2}),

\begin{equation}
V_{\rm eff}'' (r_c)= 2 \frac{ \{ y^3 A_3(\tilde{m})+y^2 A_2(\tilde{m})+y A_1(\tilde{m})+ A_0(\tilde{m}) \} }{(1-2 \tilde{m}) r_c^2 (4+y)^2 [4(1-3 \tilde{m})-y(3-5 \tilde{m})]} >0,   
\label{V2}
\end{equation}

where we have used  $\tilde{m}=m/r_c$ and $y=B^2r_c^2$ to compress the expression, and

\begin{eqnarray}
A_0(\tilde{m}) &=& 64 [- 6 \tilde{m}^2 + \tilde{m}] ,\nonumber\\
A_1(\tilde{m}) &=& 672 \tilde{m}^2 -624 \tilde{m} +128 ,\nonumber\\
A_2(\tilde{m}) &=& -200 \tilde{m}^2 +204 \tilde{m} -48 ,\nonumber\\
A_3(\tilde{m}) &=& 30 \tilde{m}^2 -37 \tilde{m} +12. 
\end{eqnarray}
The denominator in Eq. (\ref{V2}) is positive since $r > 2m$, then
the condition for stable orbits, $V_{\rm eff}^{''}  > 0$ amounts to the factor in curly brackets being positive,

\begin{equation}
 \{ y^3 A_3(\tilde{m})+y^2 A_2(\tilde{m})+y A_1(\tilde{m})+ A_0(\tilde{m}) \} > 0.   
\end{equation}

\begin{figure}[h]
\begin{center}
\includegraphics [width =0.45 \textwidth ]{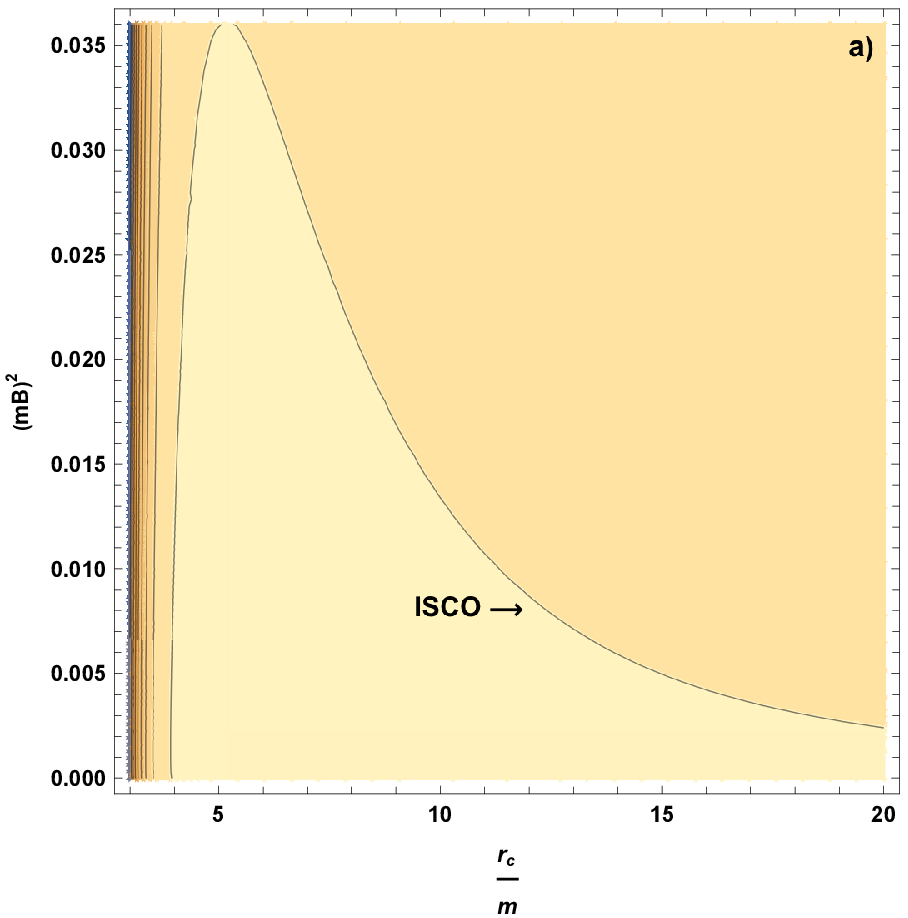}
\includegraphics [width =0.45 \textwidth ]{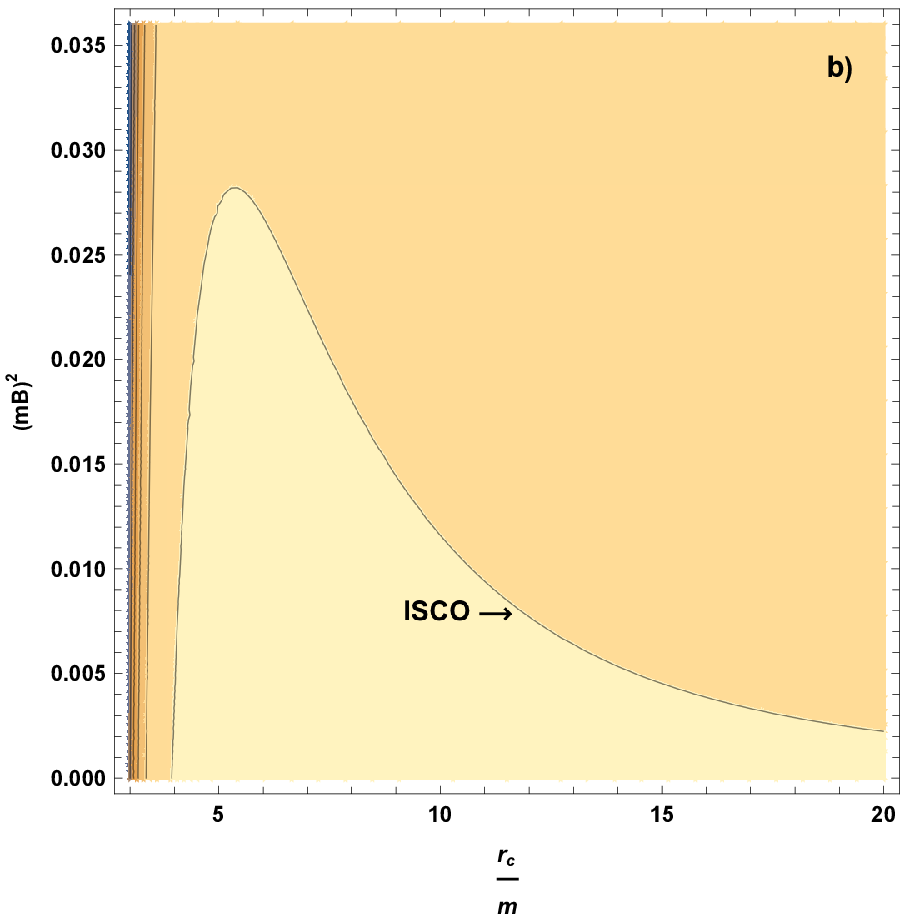}
\end{center}
\caption{\textbf{a)} The pairs ($r/m, (mB)^{2}$) corresponding to ISCO for charged test particles with $\kappa =0.4$. In \textbf{b)} The respective pairs of ISCO for uncharged test particles ($\kappa =0$) are shown. In both $L/m=6$. The regions below the ISCO curve are the ones for SCO in each case. }
\label{Fig1.1}
\end{figure}

Moreover, the condition for ISCO  is that the previous factor be zero. ISCO are the marginally stable orbits: circular orbits with radius less than the ISCO are unstable and those with radii larger than it are stable. Then the ISCO defines the border of the region of bound orbits, or the inner radius of an accretion disk.
For charged and uncharged particles, the ISCO are shown in Fig \ref{Fig1.1}, fixing the values of $\kappa$ and  $L$ and varying $r_{c}/m$ and  $(mB)^{2}$. In Fig. \ref{Fig1.1}  \textbf{a)} the ISCO  for charged particles are shown and in Fig. \ref{Fig1.1} \textbf{b)} for the uncharged particles in the same ranges of $r_{c}/m$ and  $(Bm)^{2}$.  The boundary between the  regions represents the ISCO, while the pairs ($r_c/m, (mB)^2$) corresponding to stable circular orbits (SCO) are in the region below the curve. For fixed $L$, the available region of SCO is smaller for uncharged particles.


In the next section we establish the relationship between the parameters of the stable circular orbits ( orbits obeying $V_{\rm eff}(r_c)=V_{\rm eff}^{'}(r_c)=0$ and $ {V_{\rm eff}}^{''}(r_{c})>0$) and the redshift of the light emitted by particles traveling along those geodesics.
\section{The red-blueshifts of the photons emitted by particles orbiting the Ernst BH}

The connection between the red-blueshifts of the photons emitted by massive particles (stars or gas) that move in stable geodesics around a black hole was pointed out in\cite{Herrera-Aguilar2015}. Let us start by giving a summary of the method considering a static spacetime (for more details see \cite{Becerril2016}).
\subsection{The red-blueshifts of the photons emitted by charged particles}

Light  emitted  from massive particles moving along stable circular geodesics is characterized by a $4$-momentum $\kappa^{\mu}$ that is a null vector, $\kappa^{\mu}\kappa_{\mu}=0$; photon energy and angular momentum are conserved quantities, $-E_{\gamma}=g_{tt}\kappa^{t}$  and $L_{\gamma}=g_{\phi\phi}\kappa^{\phi}$,  and the photon impact parameter is defined as ${b}={L}_{\gamma}/{E}_{\gamma}=\pm \sqrt{-{g_{\phi\phi}}/{g_{tt}}}$. 

The frequency shift ${z}$ associated with the emission (e) and detection (d) of photons emitted from particles in circular geodesics (${U}^{r}=0$) and equatorial motion (${U}^{\theta}=0$), is given by 

 \begin{equation}
1+ {z}=\frac{{U}^{t}_{e}-b_{e} {U}^{\phi}_{e}}{{U}^{t}_{d}-b_{d}{U}^{\phi}_{d}}.
\label{rdsft}
\end{equation}

Considering that observational redshifts are reported in terms of the kinematic frequency shift,  $z_{kin}= z- z_c$, where $z_c$ is the shift of a photon emitted by a static particle at $b=0$ (on the line going from the center of coordinates) and using the previous Eq.  (\ref{rdsft}), $z_c$ can be written as

\begin{equation}
1+ {z}_c=\frac{{U}^{t}_{e}}{{U}^{t}_{d}}.
\label{rdsft2}
\end{equation}
For geodesics with ${U}^{r}=0$ and ${U}^{\theta}=0$, in this same context the kinematic frequency shift  can be expressed as;

\begin{equation}\label{Zkinq}
{z}_{kin}= \frac{{U}^{t}_{e}{U}^{\phi}_{d}b_{d}-{U}^{t}_{d}{U}^{\phi}_{e}b_{e}}{{U}^{t}_{d}({U}^{t}_{d}-b_{d}{U}^{\phi}_{d})};
\end{equation}
if we consider  that the detector is located far away from the black hole then from (\ref{Lc}) and (\ref{Zkinq}) we obtain

 \begin{equation}\label{redbluecharged}
{z}={U}^{\phi}_{e}b_{+}\mid_{r_{c}}=\sqrt{-\frac{g_{\phi\phi}}{g_{tt}}}\left(\frac{{L}}{g_{\phi\phi}}-\frac{\kappa A_{\phi}}{g_{\phi\phi}} \right)\mid_{r_{c}}.
\end{equation}

Therefore the redshift ${z}$, Eq. (\ref{redbluecharged}), of the light emitted by charged particles from a stable circular orbit of radio $r_{c}$ in the equatorial plane of the Ernst BH is determined by 

\begin{equation}
{z}\mid_{r_{c}}=\sqrt{\frac{1}{r(r-2m)}}\left({L}-\kappa A_{\phi} \right).
\end{equation}

\begin{figure}[h]
\begin{center}
\includegraphics [width =0.45 \textwidth ]{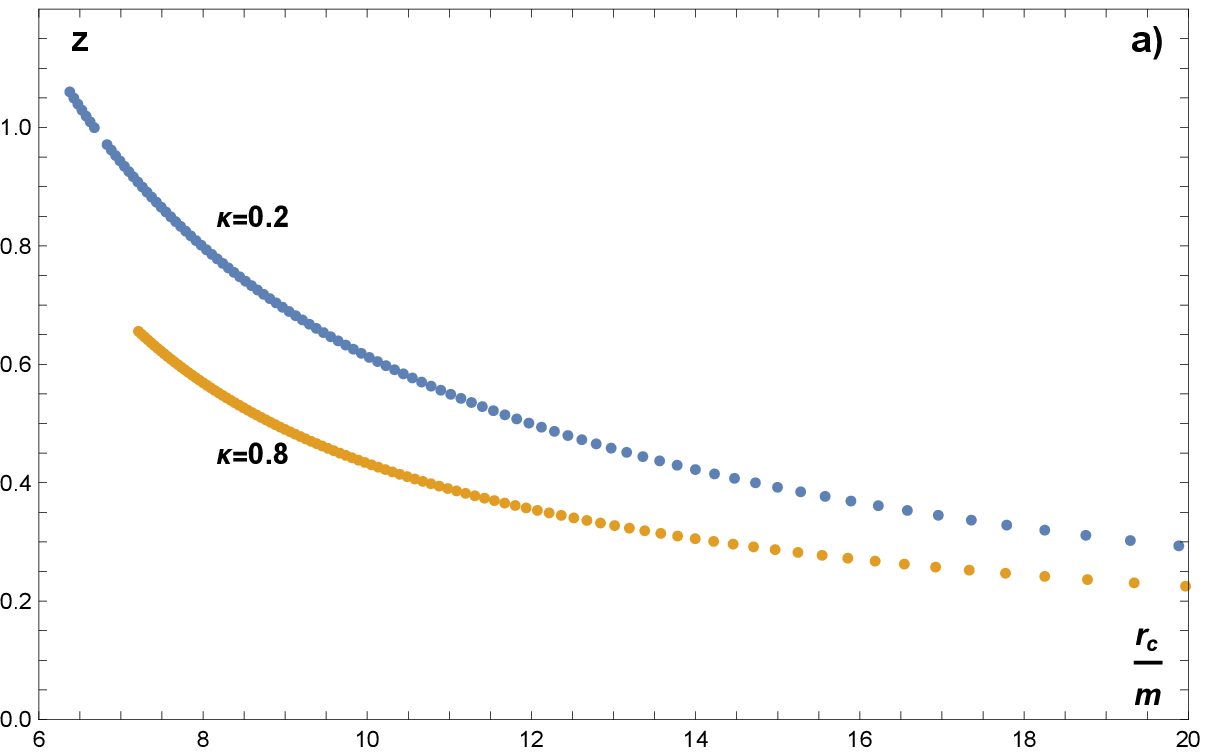}
\includegraphics [width =0.45 \textwidth ]{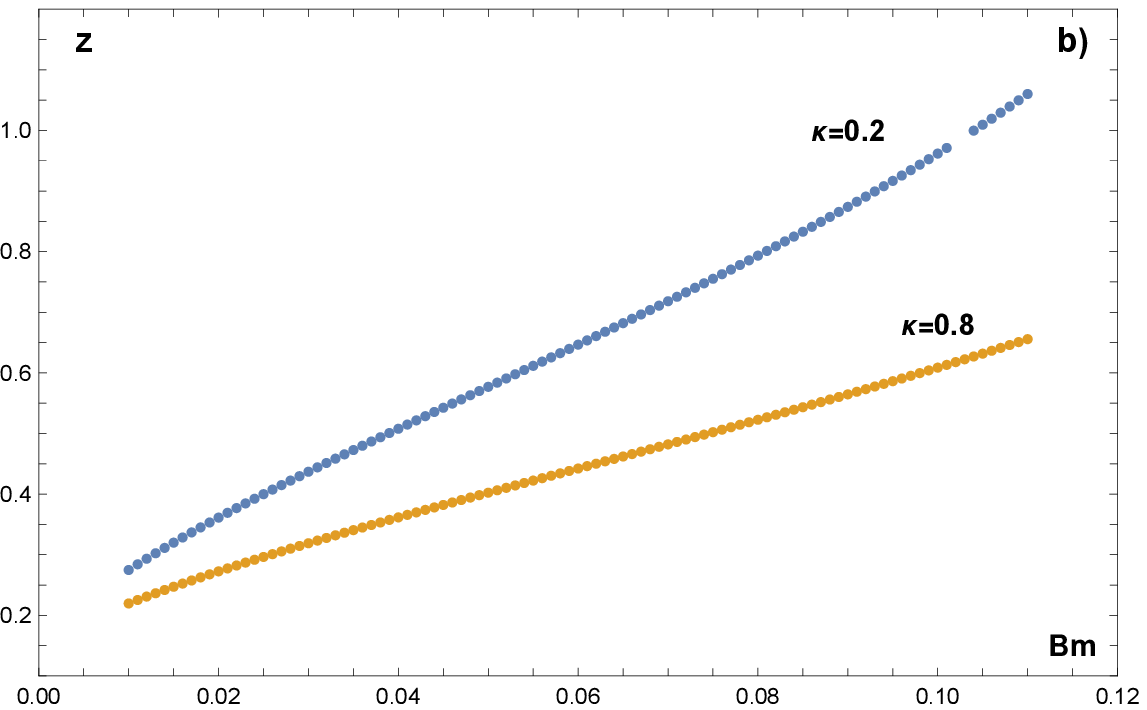}
\end{center} 
\caption{The redshift of light emitted by charged particles in stable circular orbits of the Ernst BH.
a) The redshift ${z}$ as a function of $\frac{r_{c}}{m}$,  and b) ${z}$ as a function of $Bm$; in both are fixed ${L}/m=6$.  }
\label{Fig6}
\end{figure}

Choosing  appropriate values of  ${L}$ from the ranges shown in Fig. \ref{Fig2} and  from Eq.  (\ref{Lk}) we determine the behavior of the redshift $z$ for the Ernst BH, as shown in Fig.  \ref{Fig6} for different values of $\kappa$ and varying  $mB$ in the range $(0,0.11)$.
The redshift  ${z}$ for the Ernst BH as a function of ${r_{c}}/{m}$ is shown in Fig. \ref{Fig6} \textbf{a)}; we observe that ${z}$ decreases as ${r_{c}}/{m}$  augments; the behaviour is similar for the photons emitted by neutral particles but the available range of stable orbits is smaller.
In Fig. \ref{Fig6}\textbf{b)} the behavior of ${z}$ of the Ernst BH for different values of the specific charge  $\kappa$ is shown; increasing $mB$ the redshift ${z}$ increases as well. i.e. the redshift is larger coming from particles in orbits nearer the horizon and $z$ is larger as well for a BH immersed in a magnetic field than for Schwarzschild.
\subsection{Redshift from uncharged particles orbiting the Ernst BH}

\begin{figure}[h]
\begin{center}
\includegraphics [width =0.45 \textwidth ]{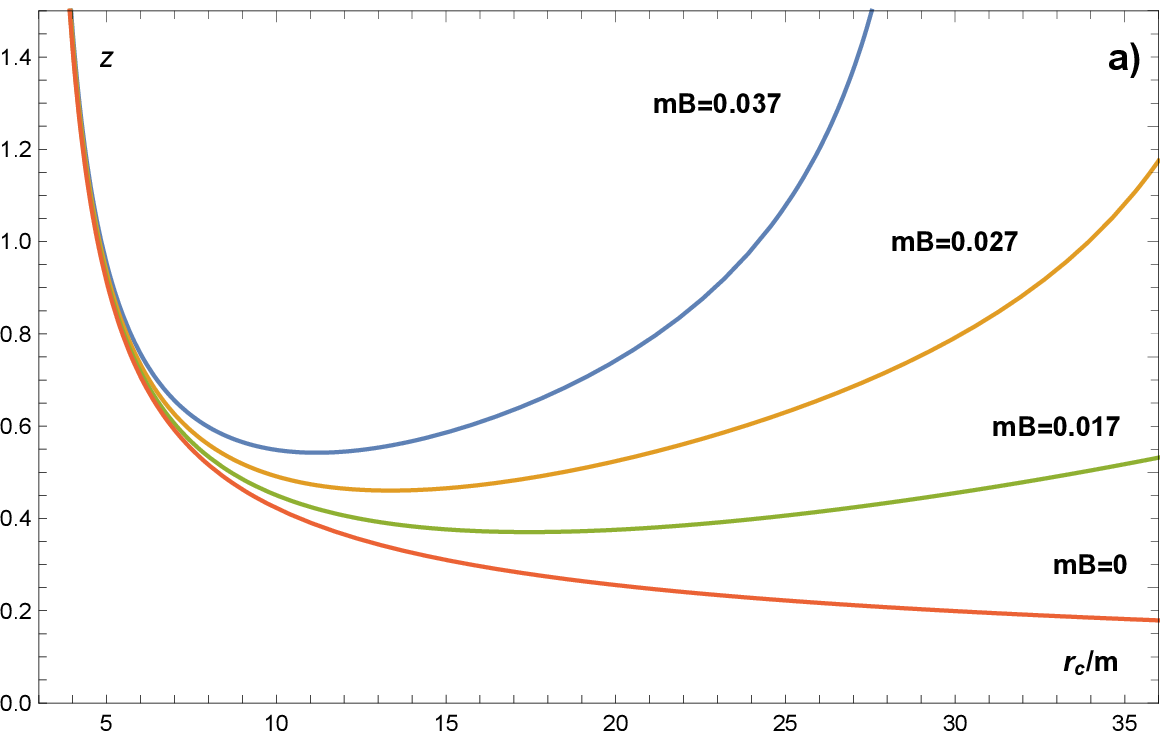}
\includegraphics [width =0.45 \textwidth ]{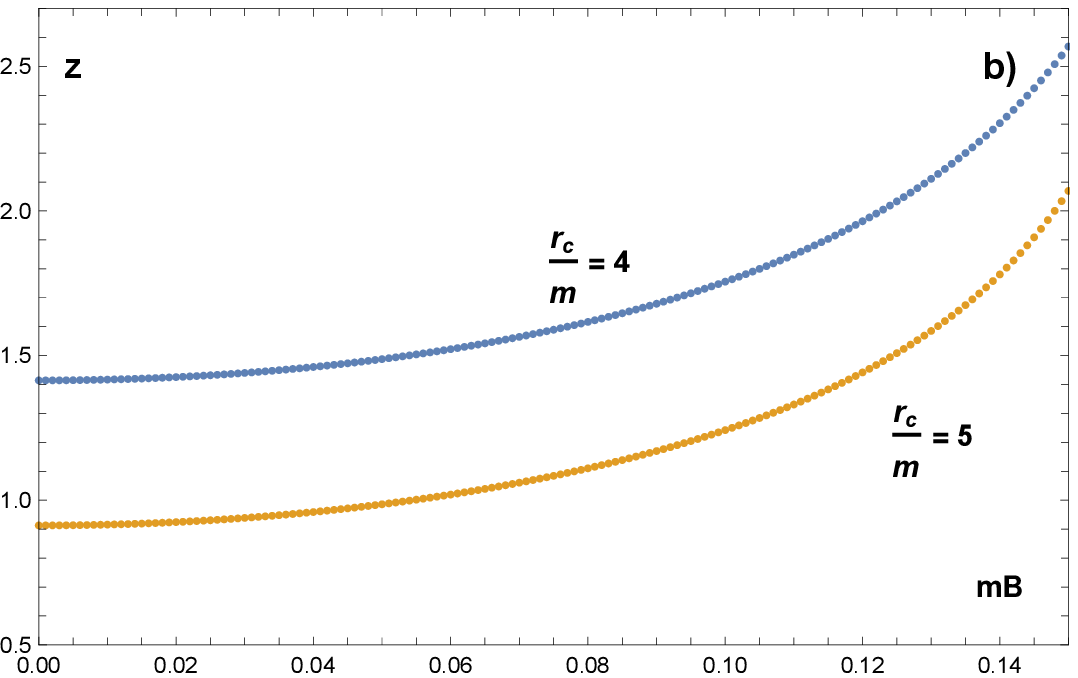}
\end{center} The redshift of light emitted by neutral particles in SCO around the Ernst BH.
\caption{a) Behavior of $z$ respect to $\frac{r_{c}}{m}$, for different values of $Bm$ and b)The redshift $z$ as a function of $Bm$, for two values of the radius of the circular orbits from which light is emitted, $\frac{r_{c}}{m}= 4, 5$. For fixed $r_c$ the presence of the magnetic field increases the redshift.}
\label{Fig3}
\end{figure}

The redshift $z$ of the light emitted by neutral particles in a stable circular orbit of radio $r_{c}$ in the equatorial plane around the Ernst BH is given by

\begin{equation}
z^2 \mid_{r_{c}}= \frac{r [m \Lambda + r \Lambda' (r- 2 m)]}{\Lambda^2 (r- 2 m)[\Lambda (r- 3m)  -2 r \Lambda' (r- 2 m)]}, 
\label{shift}
\end{equation}
since the radius is always larger than the one of the event horizon, $r>2m$, then the condition that $z^2 > 0$ reduces to $[\Lambda (r- 3m)  -2 r \Lambda' (r- 2 m)]>0$;
this condition reduces to the inequality  (\ref{condB2}) that we derived from the
requirements that $E^2 \ge 0$ and $L^2 \ge 0$. Remember that $r_c > 3m$, i.e. the radius of stable orbits is always greater than $3m$ that is the photosphere radius of Schwarzschild BH. In the case $B \rightarrow 0$ we recover the redshift $z$ for Schwarzschild BH (see \cite{Becerril2016}). 
From the previous expression we can determine $m= r_c \mathcal{G} (z^2, B^2 r_c^2)$ with

\begin{equation}
\mathcal{G} (z^2, B^2 r_c^2)= \frac{-64+48 y + (4+y)^2 (11y-20 ) z^2 \pm P(z^2,y)}{4 (4+y)^2 (5y-12) z^2},
\label{mG}
\end{equation}
where $P(z^2,y)=\sqrt{256 (4-3y)^2 -32 (4+y)^3 (7y-20) z^2 +(4+y)^6 z^2}$, with $y=B^2 r_c^2$. The redshift $z$, Eq.  (\ref{shift}), of light coming from uncharged particles is shown in Fig. \ref{Fig3} for different values of $mB$ and ${r_{c}}/{m}$. Fig. \ref{Fig3} \textbf{a)} shows  $z$ as a function of $r_{c}$; for $mB \rightarrow 0$ the behavior of $z$ is the one of Schwarzschild BH. 
The redshift $z$ for the Ernst BH decreases as $\frac{r_{c}}{m}$  augments. For values of $mB > 0.17$ the curve for $z$ presents a minimum and then increases; the ranges for $\frac{r_{c}}{m}$ are the same as in Fig.  \ref{Fig2}. It is worth to mention that the Schwarzschild BH redshift is less than the one coming from the Ernst BH $z^{S}<z^{E}$, for a given BH mass, so that  the effect of the magnetic field is of increasing the redshift but the number of stable orbits is less.
In Fig. \ref{Fig3} \textbf{b)} is displayed  $z$ as a function of $mB$ for two values of the radius of the circular orbits from which light is emitted $\frac{r_{c}}{m}= 4, 5$. For fixed $r_c$ the presence of the magnetic field increases the redshift. 

Moreover, from the condition $V_{\rm eff}''(r_c)=0$ we obtain a restriction for the ISCO, as a quadratic equation for $\tilde{r_c}= r_c/m$, 

\begin{equation}
[30 y^3-200 y^2 +672 y -384]+ [-37 y^3 +204 y^2 -624 y + 64]\tilde{r_c} + [12y^3 -48 y^2 +128 y] \tilde{r_c}^2 =0,    
\end{equation}
here $y=(mB)^2 \tilde{r_c}^2$; the previous Eq. is equivalent to a cubic equation for $y$,

\begin{equation}
y^3[30 - 37 \tilde{r_c} +12  \tilde{r_c}^2]+ y^2[-200 + 204 \tilde{r_c} -48  \tilde{r_c}^2] + y [672 - 624 \tilde{r_c} +128  \tilde{r_c}^2] + 64 \tilde{r_c}-384 =0.
\end{equation}

The 4-velocities of test particle emitters along the stable circular orbits in terms of the BH parameters $(m,B)$ and $r_c$ are

\begin{equation} 
(U^{\phi})^2= \frac{(4+y)^2[4m+y(2r_c-3m)]}{16r_c^2 [4(r_c-3m)-y(3r_c-5m)]}, \quad 
(U^{t})^2= \frac{16r_c(4-y)}{(4+y)^2 [4(r_c-3m)-y(3r_c-5m)]},
\end{equation}
that impose the restriction on the magnetic field $B$, that $(4-y)>0$ or $B^2< 4/r_c^2$.
The angular velocity of the emitters in these circular geodesics is

\begin{equation}
\Omega^2= \frac{(4+y)^4[4m+y(2r_c-3m)]}{256 r_c^3 (4-y)},
\end{equation}
recalling that $m=r_c \mathcal{G} (z^2, r_c, B^2)$, Eq. (\ref{mG}), 
indeed these velocities, $U^{\phi}, U^{t}, \Omega$ depend on the magnetic field and
correspond to a certain redshift $z$. In such a way that given a set of observables $\{ z, r_c \}_{i}$, Bayesian statistical analysis would provide an estimate for both parameters $m$ and $B$.

\section{Conclusions}

In this work we analyze  the redshift of the photons emitted by massive and charged test particles that move around a black hole immersed in an external magnetic field, situation represented by the Ernst metric. We consider that the emitters of light are moving along stable circular orbits (SCO). 
In terms of the effective potential SCO obey that ${V_{\rm eff}}^{'}(r_{c}) = {V_{\rm eff}}(r_{c})=0$ and ${V_{\rm eff}}^{''}(r_{c}) > 0$.

The radii of the circular orbits are as well modified in presence of the magnetic field; these radii $r_c$ are shifted towards the horizon in proportion to the magnitude of the magnetic field, i.e. the minima of the effective potential $r_c$ is nearer the horizon as $mB$ augments; or $r_c^{E} > r_c^{S}$, i.e.  the radius SCO in Schwarzschild BH is less than radius of SCO for the Ernst BH. We obtain numerically the density  regions of  the pairs angular momentum-Magnetic fields $(L,B)$  that correspond to SCO of the charged test particles. The set depends on the specific charge of the test particle.  

Moreover, we determine the ranges of the parameters that allow the existence of circular orbits of neutral test particles corresponding to SCO. It is also presented the region  for the innermost stable circular orbits (ISCO) in terms of the radius of the orbit $r_c$ and the dimensionless parameter $mB$. We found that there is an upper bound for the magnetic field that allow SCO, $(mB)_{\rm max}=0.189366$; this field could correspond to a ten solar masses BH with a magnetic field of $B_{\rm SI} \approx 10^{16}$, or to a lesser magnetic field with a more massive BH (see table \ref{table1}). 

We have shown that the magnetic field affects, through the curvature,  neutral or uncharged test particles: the redshift emitted in the presence of the magnetic field is larger than the shift in absence of the field, i. e. light coming from particles orbiting the Schwarzschild BH is less redshifted than the one coming from the BH immersed in a magnetic field.

In summary,  the presence of the magnetic field in the vicinity of the BH, considering SCO of neutral and charged particles, enlarges the redshift of the light coming from test particles orbiting the BH. This effect should be taken into account in observations, mainly when strong magnetic field are involved; otherwise it could lead to overestimate the BH mass.

\section*{ACKNOWLEDGEMENTS}

\vspace{0.5cm}
\textbf{Acknowledgments}:  N. B. acknowledges partial financial support from CONACYT-Mexico through the project No. 284489. The authors acknowledge  financial support from SNI-CONACYT, Mexico.



\begin{thebibliography}{9}

\bibitem{Abbott2016}
B. P. Abbott et al, {\it Observation of Gravitational Waves from a Binary Black Hole Merger},
Phys. Rev. Lett., {\bf 116} (6) 061102 (2016).


\bibitem{2009ASSP}
F. Eisenhauer et al, {\it GRAVITY: Microarcsecond Astrometry and Deep Interferometric Imaging with the VLT},
Astr. Space Sc. Proc. {\bf 9}, 361 (2009).

\bibitem{Akiyama2019} 
K. Akiyama et al, {\it First M87 Event Horizon Telescope Results. I. The Shadow of the Supermassive Black Hole},
Astrophys. J. {\bf 875} (1), L1 (2019).
  
\bibitem{Cardoso2009}
V. Cardoso, A. S. Miranda, E. Berti, H. Witek, V. T. Zanchin,
{\it Geodesic stability, Lyapunov exponents and quasinormal modes},
Phys. Rev. D {\bf 79}, 064016 (2009).
     
\bibitem{Fernando2012}
S. Fernando, J. Correa,
{\it Quasinormal Modes of Bardeen Black Hole: Scalar Perturbations},
Phys. Rev. D {\bf 86} 064039 (2012).
     
\bibitem{Breton2016}
N. Breton, L. A. Lopez, 
{\it Quasinormal modes of nonlinear electromagnetic black holes from unstable null geodesics},
Phys. Rev. D {\bf 94} (10) 104008 (2016).

\bibitem{Lopez2018}
L. A. Lopez, V. Hinojosa,
{\it Quasinormal modes of Charged Regular Black Hole},
Can J. Phys. {\bf 99} (1) 44-48 (2021).

\bibitem{Konoplya2017}
R.A. Konoplya, Z. Stuchlík, {\it Are eikonal quasinormal modes linked to
the unstable circular null geodesics? }
Phys.Lett. B {\bf 771} 597-602 (2017)

\bibitem{Herrera-Aguilar2015}
A. Herrera-Aguilar, U. Nucamendi, 
{\it Kerr black hole parameters in terms of the redshift/blueshift of photons emitted by geodesic particles},
Phys. Rev. D {\bf 92} (4) 045024 (2015).

\bibitem{Becerril2016}
R. Becerril, S. Valdez-Alvarado, U. Nucamendi, 
{\it Obtaining mass parameters of compact objects from redshifts and blueshifts emitted by geodesic particles around them},
Phys. Rev. D {\bf 94} (12) 124024 (2016).

\bibitem{Kraniotis2021}
G. V. Kraniotis,
{\it Gravitational redshift/blueshift of light emitted by geodesic test
particles, frame-dragging and pericentre-shift effects, in the
Kerr–Newman–de Sitter and Kerr–Newman black hole geometries},
Eur. Phys. J. C {\bf 81} 147 (2021)

\bibitem{Begelman2003}
M. C. Begelman, {\it Evidence for Black Holes},
Science, {\bf 300}, 1898-1903 (2003).

\bibitem{Ernst1976}
F. Ernst, {\it Black holes in a magnetic universe},
J. Math. Phys. {\bf 17}, 54-56 (1976).

\bibitem{Hoenselaers1979}
N. Dadhicht, C. Hoenselaers and C. V. Vishveshwara,
{\it Trajectories of charged particles in the static Ernst space-time}, J. Phys. A: Math. Gen. {\bf 12} 215-221 (1979).

\bibitem{Wald1974}
R. M. Wald, 
{\it Black hole in a uniform magnetic field},
Phys. Rev. D, {\bf 10} (6), 1680-1685 (1974).

\bibitem{Frolov2010}
V. P. Frolov, A. A. Shoom, {\it Motion of charged particles near a weakly magnetized Schwarzschild black hole},
Phys. Rev. D, {\bf 82}, 084034 (2010).

\bibitem{Lim2015}
Y. K. Lim, {\it Motion of charged particles around a magnetized/electrified black hole},
Phys. Rev. D, {\bf 91} 024048 (2015).

\bibitem{Tursonov2016}
A. Tursunov, Z. Stuchlík,  M. Kolos,
{ \it Circular orbits and related quasi-harmonic oscillatory motion of charged particles around weakly magnetized rotating black holes}, Phys. Rev. D {\bf 93} 084012 (2016).

\bibitem{Hackmann2020}
J. P. Hackstein, E. Hackmann, {\it Influence of weak electromagnetic fields on charged
particle ISCOs},  Gen. Relativ. and Grav. {\bf 52}, 22,  (2020).

\bibitem{Aliev2002}
A. N. Aliev, N. \"Ozdemir, {\it Motion of charged particles around a rotating black hole in a magnetic field},
M. N. R. A. S. {\bf 336}, 241-248 (2002).

\bibitem{Konoplya2007}
R.A. Konoplya, {\it  Magnetized black hole as a gravitational lens},
Phys.Lett. B {\bf 644} (2007) 219-223

\bibitem{Walter1980} 
W. J. Walter, R. M. Kerns, {\it Surface geometry of a black hole in a magnetic field},
Phys. Rev. D, {\bf 21} 332-335 (1980).

\bibitem{Melvin1965}
M. A. Melvin, {\it Dynamics of Cylindrical Electromagnetic Universes},
Phys.  Rev. {\bf 139B} 225-243 (1965).

\bibitem{Konoplya2008}
R. A. Konoplya, R. D. B. Fontana, {\it Quasinormal modes of black holes immersed in a strong magnetic field},
Phys. Lett. B {\bf 659}, 375-379 (2008).

\bibitem{Piotr2010}
M. Yu. Piotrovich, N. A. Silant\'ev, Yu. N. Gnedin, T. M. Natsvlishvili, 
{\it Magnetic Fields of Black Holes and the Variability Plane},
arXiv:1002.4948.


\end{thebibliography}

\end{document}